\documentclass{aa}

\usepackage{graphicx}
\usepackage{txfonts}
\begin{document}

   \title{Magnetic fields in the accretion disks for various inner boundary conditions}

   \author{D.V. Boneva
          \inst{1}
          \and
          E.A.Mikhailov\inst{2,3,4}
          \and
          M.V.Pashentseva\inst{2}
          \and
          D.D.Sokoloff\inst{2,3,5}
          }

   \institute{Space Research and Technology Institute, Bulgarian Academy of Sciences,
              Acad G. Bonchev St., Bl. 1, 1113 Sofia, Bulgaria\\
              \email{danvasan@space.bas.bg }
         \and
             M.V. Lomonosov Moscow State University, Leninskie gori, Bl.1, 119991 Moscow, Russia\\
              \email{ea.mikhajlov@physics.msu.ru}
         \and
            Moscow Center of Fundamental and Applied Mathematics, Leninskie gori, Bl.1, 119991 Moscow, Russia\\
         \and
            P. N. Lebedev Physical Institute, Leninskiy Prospekt, Bl.53, 119333 Moscow, Russia\\
         \and
             N.V. Pushkov Institute of Terrestrial Magnetism, Ionosphere and Radio Wave Propagation (IZMIRAN), Kaluzhskoe Hwy., Bl.4, 108840, Troitsk, Moscow, Russia\\
             \email{sololoff.dd@gmail.com}
                          }

   \date{Received ...; accepted ...}


  \abstract
   {The magnetic fields of accretion disks play an important role in studying their evolution. We may assume that its generation is connected to the dynamo mechanism, which is similar with that in the galactic disks.}
   {Here, we propose a model of the magnetic field of the accretion disk that uses the same approaches that have been used for galaxies. It is necessary to obtain the field, which is expected to be less than the equipartition value, and without destroying the disk. To do so, it is necessary to formulate the basic properties of the ionized medium and to estimate the  parameters  governing the dynamo.}
   {We used the no-$z$ approximation that has been developed for thin disks. We also take different boundary conditions that can change the value of the field significantly.}
   {We show that the magnetic field strictly depends on the boundary conditions. Taking zero conditions and the fixed magnetic field condition on the inner boundary, which are connected to the physical properties of the accretion disk, we can avoid solutions that are greater than the equipartition field.}
   {}


   \maketitle
%

\section{Introduction}

   The accretion disks surrounding various objects, such as neutron stars, black holes, or white dwarfs, play an important role in astrophysical processes ~\citep{Shakura73,LyndenBell74,Tylenda81,Horne86,Lin87,Gansicke06,Jiang19}. Nowadays, there is no doubt that magnetic fields take part in different processes in such disks, with the capacity to explain various phenomena. The magnetic field activity can explain the processes of angular momentum transport from one part of the disk to another~\citep{Shakura73}. In disks surrounding protostellar objects, we can assume hydromagnetic winds, which are connected to magnetized molecular disks ~\citep{Pudritz86}. They  are also connected to the jet production and structure in different objects~\citep{Balbus91,Blandford82,Lovelace14}. It is important to understand and research the mechanisms of the magnetic field generation and the evolution of the turbulence in the medium in the accretion disk. This could help us in further studies of related  processes.

   There are different parts of this problem. In relation to the magnetic field, it can be connected to the transport of the accreting matter ~\citep{Lubow94,Okuzumi14}. In turn, the occurrence of turbulence is an equally difficult issue, which leads to even more confusion. It may be connected  to the accreting matter, but different authors ~\citep{Brandenburg97} reasonably describe it as the result of the magneto-rotational instability. However, the turbulence is the theme of a specific research, and we are only trying to describe the evolution of large-scale magnetic field structures. When the turbulence is already generated, we can solve the mean field equations using different methods.

  In previous works, the magnetic field in the disk is described by different approaches. Excluding the transport of the accreting medium~\citep{Lubow94}, it is necessary to mention the possibility of connecting the magnetic field in the disk with the one in the central object of the accreting system. But the observational data and theoretical models of the accretion disks give us an opportunity to assert that we may neglect the interplay between the disk and the central body, in most cases, and to explain the magnetic field generation via the dynamo mechanism ~\citep{Stepinsky90,Torkelsson94,Rudiger95,ReyesRuiz95}. Certainly, for such objects as X-ray pulsars, the interaction may be important, but this ought to be the subject of a specific study.

  The basic principles of the dynamo mechanism are connected to the transition of kinetic energy of turbulent motions to the energy of the magnetic field~\citep{Parker55,Moffatt78,Krause80,Molchanov85,Sokoloff15}. The field generation is usually based on two main effects. Firstly, the celestial bodies (or their parts) rotate with changing angular velocities that are connected to the rotation and stretching of the magnetic field. Secondly, there are turbulent motions and their vorticity describes the alpha-effect, which is connected to the rotation of the field line. The joint action of these two effects cause the field amplification, which competes with the dissipation effects that destroy the regular structures of the magnetic field. Thus, the dynamo is a threshold process: the magnetic field can be generated only if the alpha-effect and differential rotation are more intensive than the dissipation effects. In this case, it grows according to an exponential law. Otherwise, the magnetic field decays. There is a wide spectrum of different models that have reduced the general magnetohydrodynamics equations to systems characterizing the magnetic field of specific objects~\citep{Zeldovich83,Brandenburg18}.

   As the results of many studies have shown, the accretion disk structure and the magnetic field generation process are mostly similar when the structures of galaxies are similar. The velocity of large-scale motions changes depending on the distance from the center (it is the base of differential rotation), and as for the turbulent motions, we have the mirror asymmetry, which gives us the opportunity to speak about an alpha-effect. Thus, we can use approaches of the disk dynamo ~\citep{Rekowski04} that have been developed for the galaxies.

   Of course, it is possible to use direct numerical simulations ~\citep{Stone94,Stone96,Pariev07,Davis10} to find the magnetic field, but the results of such studies describe the field structure, which is, in principle, quite similar to that of the galaxy ~\citep{Jiang13}. This is expected because both the accreting systems and the galaxies can be presented as thin disks ~\citep{Gissinger09}, where the typical kinematic processes are nearly the same. However, it is important to study the structure of this magnetic field in detail, and to do so, we can take main approximations that have been established for the galaxies.

   The galactic dynamo is often described by the so-called "no-$z$ approximation," which has been developed for thin disks ~\citep{Moss95}. It makes use of the fact that the main part of the magnetic field lies in the equatorial plane. So, we can solve the equations only for radial and azimuthal components of the field. The vertical component can be taken from the solenoidality condition. Also, we can change some of the partial derivatives of the magnetic field by algebraic expressions, and the equations become much simpler both for the theoretical estimates and the numerical study. This approach was used in the work of Moss et al. (2016) The magnetic field of the accretion disks surrounding white dwarfs was studied using the no-z approximation. It is necessary to mention that the turbulence and the vorticity of the motions in accretion disks are connected to magnetorotational instability; and in the galaxies, the main source of turbulence comes from the supernovae explosions ~\citep{Brandenburg97}. Fortunately, in such models, the turbulence is assumed to be given and is characterized by the dimensionless parameter that does not correspond to its origin. The results obtained in this approximation~\citep{Moss16} show that the field has a structure that is quite close to the models based on the results of direct numerical simulation of such objects ~\citep{Brandenburg95}.

Let us discuss the idea of our research from the viewpoint of galactic dynamo studies. It is appealing to believe that there is some recognizable similarity between the problems. Galactic dynamo modeling allows for the development of simple pragmatic models that do not pretend to address a deep understanding of the process, but that allow us to obtain robust and realistic radial magnetic field distribution. The possibility of such a model is, in particular, based on the fact that the strength of excited large-scale magnetic field is comfortably lower than the equipartition with turbulent motion. We appreciate that a dynamo can generate a large-scale magnetic field with magnetic energy that exceeds equipartition with kinetic energy of turbulent motions, and we obtained such results to support this in our previous works (see, e.g.,~\citep{Moss16}. However, it is quite interesting to move on and study whether the field can be smaller than the equipartition value. Here, we demonstrate that it can be done using the model that has its main feature tied to another boundary condition.

Also, higher values of the magnetic field indicate that the generation of such a strong magnetic field deserves a fully nonlinear dynamo model that includes a detailed description of magnetic force. Such forces can influence the structure of the turbulent motions, for example, by changing its spectra~\citep{Mikhailov19}. For some specific cases, when there is an imposed magnetic field onto the accretor itself, it can even cause deformations in the accretion disk~\citep{Rudiger02}.

In this work we consider a quite qualitative model that uses the given parameters of turbulence in the disk. Certainly, there are models that describe the field evolution based on magnetorotational instability without the implementation of any other form of turbulence~\citep{Hawley95,Matsumoto95,Brandenburg95}. For this case, the turbulent velocity is much smaller than other typical velocities and the dynamo number can be assumed to be very large. While the magnetic field grows, the turbulence becomes more intensive because of the magnetorotational instability, and it leads to saturation of the growth.

   While studying the magnetic field, we pay special attention to the boundary condition, which describes the field at the inner boundary of the disk. However, this condition describes the interaction between the disk and the central body, so it strongly influences the field generation process. We tried to mimic this effect and as a result, the magnetic field becomes quite modest and lower than the equipartition value. Also, we should emphasize that it is necessary to accurately describe different parameters of the disk.

   The paper is organized as follows. In Sect.2, we describe the main properties of the disk in connection with the basic drivers of the dynamo mechanism (such as the turbulent velocity, angular velocity, and the geometrical parameters of the disks), indicating the boundary conditions. In Sect.3, we present the dynamo model in its application to the accretion disk, and we give the typical solutions for the magnetic field in different cases (we discuss the boundary conditions and another parameters of the model). We show in Sect. 4 that the clarification of the model can give us the magnetic field, in good accordance with basic principles and does not exceed the equipartition value.

\section{Properties of the medium in the accretion disks}

   To describe the dynamo mechanism in accretion disks, it is necessary to know the basic properties of the medium and some disk parameters. Here, we point only to those that are related to our calculations of binary stars with accretion disk, considering the thin disk approximation.

   In our assumption, the matter moves throughout the disk in a circular orbit. Its angular velocity $\Omega$ has the Keplerian value and it could be expressed by:
   \begin{equation}\Omega =\sqrt{\frac{GM}{r^{3}}},\end{equation}
   where $M$ is the mass of the central object or the mass of the primary star, using the terminology of binary stars configuration, and $r$ is the cylindrical polar coordinate corresponding to distance from the central object ~\citep{Frank2002}. The angular velocity is assumed to be highly supersonic.

   In the direction of the accretor, as a result of the dissipative processes, the matter moves also radially with radial velocity $V_{r}$:
  \begin{equation}V_{r}\approx \alpha_{t} \left(\frac{H}{R_d}\right)^{2} V_{\varphi},\end{equation}
   (Lipunov 1982). For the circular velocity, we can write: $V_{\varphi}=r \Omega_k(r)$  and $H$ is the disk half-thickness, $R_d$ is the radius of the accretion disk, $\frac{H}{R_d} - $  the ratio of the disk thickness to its radius, and $\alpha_{t}$ is the turbulence parameter (usually $0.01<\alpha<1$). In the case of a steady thin disk approximation, for the radial velocity we can write \citep{Frank2002}:
   \begin{equation}V_{r}=-\frac{3 \nu}{2r} \frac{1}{1-\sqrt{\frac{r_{c.o.}}{r}}},\end{equation}
   where $r_{c.o.}$ is the radius of the central body (which is few times smaller than inner radius of the accretion disk $r_{in}$), $\nu = \alpha_{t} c_s H $ is the kinematic viscosity, $c_s$ - the speed of sound. Since $V_{r}$ is highly subsonic, which is in order of $\frac{\nu}{r}$, that is, $V_{r}\sim\frac{\nu}{r}\sim\alpha_{t} c_s\frac{H}{r}$.   If we accept the standard model on the structure of the steady disk, it can be assumed that $V_{r} \sim 10^{-1} \mbox{ km s}^{-1}\sim 10^{2} \mbox{ m s}^{-1}.$

   For the disk half-thickness, we can take the following model~\citep{Suleimanov07, Moss16}:
  \begin{equation}h(r)=h_{0}\left(\frac{r}{R}\right)^{\frac{9}{8}}\left(1-\sqrt{\frac{r_{c.o.}}{r}}\right)^{\frac{3}{20}},\end{equation}
   where $h_{0}$ is some typical value of the half-thickness, associated with the outer parts of the object.

   \section{Dynamo mechanism}

   Concerning disk dynamos, studies of the galactic magnetic field evolution have shown that the possibility of the magnetic field generation can be described by dimensionless number ~\citep{Arshakian09}:
   \begin{equation}D=\left(\frac{3 h \Omega}{v_{t}} \right)^{2},\end{equation}
   where $v_{t}$ is the turbulent velocity in the disk. The magnetic field can grow if $D>D_{cr},$ where $D_{cr} \approx 7$ is the critical dynamo number, or else the field decays.

   We can use this formula to make simple estimates and to check if the dynamo mechanism can generate the regular field in the concrete galactic or accretion disk. For example, with regard to the accretion disk near the white dwarf, we can take the mass of the central body $M=1M_{\odot} \approx 2 \cdot 10^{33} \mbox{ g}=2 \cdot 10^{30} \mbox{ kg}.$ For the outer disk radius, we can take outer radius $R=5 \cdot 10^{9} \mbox{ cm}=5 \cdot 10^{7} \mbox{ m}.$ So, the typical angular velocity is $\Omega=\sqrt{GM/R^3} \approx 0.03 \mbox{ s}^{-1}.$ For the half-thickness of the disk it is convenient to take $h=10^{-2}R,$ and the turbulent velocity of the interstellar medium is $v_{t}=15 \mbox{ km s}^{-1}=$ $=1.5 \cdot 10^{4} \mbox{ m s}^{-1},$
   values $h=5 \cdot 10^{7} \mbox{ cm}=5 \cdot 10^{5} \mbox{ m},$  $\Omega=0.033 \mbox{ s}^{-1}$. Thus, for the dynamo number, we have a value $D\approx 10,$ which is higher than the critical one. This value of the dynamo number is comparable with the one used by~\citep{Moss16}. However, for our further modeling, it is also useful to take the models that take smaller turbulent velocities, which provide larger dynamo numbers~\citep{Torkelsson94,Camenzind90}.

   \begin{figure*}
   \centering
   \includegraphics[width=11cm]{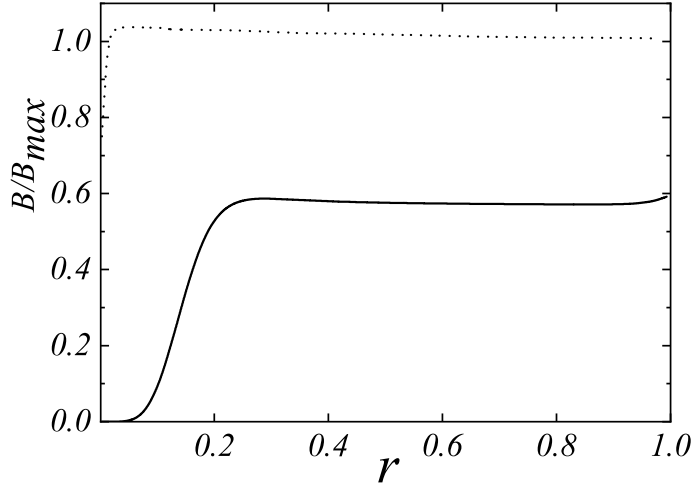}
   \caption{Dependence of the magnetic field on the distance to the center of the accretion disk for different cases ($D_{eff}=9$). Dotted line shows the results from the work of Moss et al. 2016. Solid line shows the zero condition.}
              \label{FigGam}%
    \end{figure*}

   The dynamo effect is connected to the transition of the turbulent motions energy to the energy of the magnetic field. Thus, the field growth is restricted by the equipartition value $B^{*}$, which corresponds to the case of equal densities of the kinetic energy and the magnetic field~\citep{Shukurov04}:
   \begin{equation}\frac{B^{*2}}{8 \pi}=\frac{\rho v_{t}^{2}}{2}.\end{equation}
   For the equipartition field, we have
   $B^{*}=2v\sqrt{\pi \rho}.$
  In the equations of the field evolution, it can be taken into account by nonlinear saturation: the dynamo is expected to become less intensive if the field becomes closer to the equipartition value, which can be described by multiplying the $\alpha$-effect coefficient on the $\left(1-\frac{B}{B^{*2}}\right)$

   To model the magnetic field in thin disks, we can take the no-$z$ approximation  developed by~\citep{Moss95}, along with some clarifications that have been described, for example by~\citep{Phillips01}. It is assumed that the field mainly lies in the equatorial plane, and the components of the field are proportional to $\cos\left( \frac{\pi z}{2h}\right).$ Thus, $z$-derivatives can be altered by the following algebraic expressions:
   \begin{equation}\frac{\partial^{2} B_{r, \varphi}}{\partial z^{2}}=-\frac{\pi^2}{4}B_{r,\varphi}.\end{equation}

   In the simplest case, the equations of this model can be written as (the factor $\frac{\pi^{2}}{4}$ is connected to the $z$-dependence which has been described above):
   \begin{equation}\label{eqnoz1}\frac{\partial B_{r}}{\partial t}=-R_{\alpha}B_{\varphi} \left(1-\frac{B^{2}}{B^{*2}}\right)-\frac{\pi^{2}}{4}B_{r}+\lambda^{2} \frac{\partial}{\partial r} \left(\frac{1}{r}\frac{\partial}{\partial r} \left(rB_{r}\right)\right)-\frac{V_{r}}{r}\frac{\partial (r B_{r})}{\partial r},\end{equation}
   \begin{equation} \label{eqnoz2} \frac{\partial B_{\varphi}}{\partial t}=-R_{\omega}B_{r} -\frac{\pi^{2}}{4}B_{\varphi}+\lambda^{2} \frac{\partial}{\partial r} \left(\frac{1}{r}\frac{\partial}{\partial r} \left(rB_{\varphi}\right)\right)-\frac{\partial}{\partial r}\left(V_{r} B_{\varphi} \right).\end{equation}
   Here, the distances are measured in the outer radius of the accretion disk $R,$ time is measured in units of $h^{2}(R)/\nu$ ($\nu$ is the turbulent diffusion coefficient), and the magnetic field is measured in equipartition units. The governing parameters lead us to the constant value of the dynamo number. It is necessary to note that these equations are not the result of a strict mathematical derivation, but they are formulated to be as similar as possible to those of~\citep{Moss16}. Lower down, we show the field evolution for changing dynamo number.

 \begin{figure*}
   \centering
   \includegraphics[width=11cm]{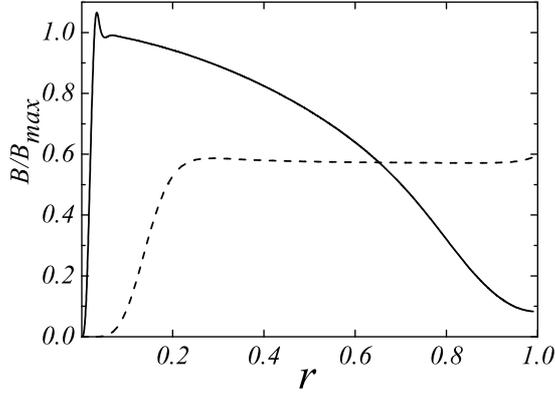}
   \caption{Dependence of the magnetic field on the distance to the center of the accretion disk for different cases. Dashed line shows constant value of dynamo number ($D=9$), solid line shows dynamo number, which depends on the distance to the center of the accretion disk ($D_{eff}=9$).}
              \label{FigGam}%
    \end{figure*}

   In these equations, some dimensionless parameters have been introduced. Also, $R_{\alpha}$ characterizes the alpha-effect (vorticity of the turbulent motions):
   \begin{equation}R_{\alpha}=\frac{\Omega(R)l^{2}}{\nu},\end{equation}
   where $l$ is the typical length scale of the turbulence. Such law is connected to the usual model for alpha-effect coefficient, which is proportional to the angular velocity of the disk and the squared turbulence length scale~\citep{Arshakian09}.

   $R_{\omega}$ describes the differential rotation:
   \begin{equation}R_{\omega}=\frac{\Omega(R)h(R)^{2}}{\nu}.\end{equation}
   $\lambda$ shows the half-thickness of the accretion disk and the role of the dissipation in the disk plane:
   \begin{equation}\lambda=\frac{h}{R}.\end{equation}

   Here, the dynamo-number will be constant and it can be calculated as product of alpha-effect and differential rotation coefficients:
   \begin{equation}D=R_{\alpha}R_{\omega}.\end{equation}

   We can take the following values for the equipartition field and the $\lambda$-coefficient \citep{Suleimanov07}:
   \begin{equation}B^{*}=420 r^{-\frac{21}{16}}\left(1-\sqrt{\frac{r_{c.o.}}{r}}\right)^{\frac{17}{40}},\end{equation}
   \begin{equation}\lambda=0.037 r^{\frac{1}{8}}\left(1-\sqrt{\frac{r_{c.o.}}{r}}\right)^{\frac{3}{20}},\end{equation}
   As for the velocity, we can use the formula:
   \begin{equation}V_{r}=\frac{V_{0}}{r\left({1-\sqrt{\frac{r_{c.o.}}{r}}}\right)},\end{equation}
 where $V_{0}$ is connected to the viscosity and the nature of the accretion disk we are  considering. For the simplest case, we can take it as constant, but for the clarification, the $r$-dependence may be introduced as well.

   To model the magnetic field, we should describe different boundary conditions. Regarding the external boundary, they describe the zero derivative for the components of the magnetic field:
   \begin{equation}\frac{\partial B_r}{\partial r}\left ( R \right ) = 0, \frac{\partial B_{\varphi}}{\partial r}\left ( R \right ) = 0.\end{equation}

   The magnetic field at the inner boundary is much more interesting. Firstly, we can describe the zero-field condition, which gives no interaction between the central body and the accretion disk:
   \begin{equation} B_r(r_{in})=0, B_{\varphi}(r_{in})=0.\end{equation}
   For this case, we can take the initial condition as
    \begin{equation}B_{\varphi}(t=0) = B_0 \sin(\frac{\pi (r - r_{in})}{(R-r_{in})}),\end{equation}
    where $B_{0}$ is some small value, which is much smaller than the equipartition field (making it quite convenient to take $B_{0}=10^{-3}$). We also tried to take the conditions of the zero field derivative at the inner boundary, but the final results are nearly the same for this case.

       \begin{figure*}
   \centering
   \includegraphics[width=11cm]{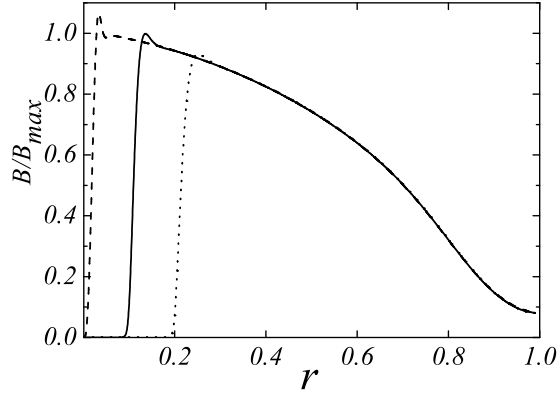}
   \caption{The dependence of the magnetic field on the distance to the center of the accretion disk for different $r_{in}$ with zero boundary condition. Dashed line shows the case $r_{in}=0.01,$ solid line -- the case $r_{in}=0.1$ and the dotted line -- the case $r_{in}=0.2$.}
              \label{FigGam}%
    \end{figure*}

    We considered the ratio $B/B^{*}$ because in similar models~\citep{Phillips01} that use no-$z$ approximations, it is usually  smaller than 1, so we can expect that it will have nearly the same values. The results for it are shown on Fig.1. Here, $D=9$, for the inner radius of the disk we take $r_{in}=0.01$ and for the radius of the central object, we use $r_{c.o.}=0.003$ (distances are measured in units of outer radius of the accretion disk). For comparison, we also give the result of the field, which has been obtained by~\citep{Moss16} for specific boundary conditions connected to the second derivative of the field component, $B_{r}$ and $\left(B/B^{*}\right)^{2}$, ratio on the inner boundary; in addition, the model was a bit different. We can see that the zero-field boundary condition  gives the field which is smaller than the equipartition field. The field does not even reach the values comparable with it and it does not accumulate near $r=r_{in},$ as shown in the work of~\citep{Moss16}. Certainly, this can be reached by using other model assumptions, but we give one possible explanation that appears quite reasonable.

    In the accretion disks, the angular velocity strongly depends on the distance from the center, so it would be effective to change the stand-alone $R_{\omega}$ coefficient by $R_{\omega}r\frac{d\Omega}{dr}.$ Then, if we measure the angular velocity in dimensionless units, it will lead us to the following equations:
    \begin{equation}\frac{\partial B_{r}}{\partial t}=-R_{\alpha}B_{\varphi} \left(1-\frac{B^{2}}{B^{*2}}\right)-\frac{\pi^{2}}{4}B_{r}+\lambda^{2} \frac{\partial}{\partial r} \left(\frac{1}{r}\frac{\partial}{\partial r} \left(rB_{r}\right)\right)-\frac{V_{r}}{r}\frac{\partial (r B_{r})}{\partial r},\end{equation}
   \begin{equation}\frac{\partial B_{\varphi}}{\partial t}=-\frac{3 R_{\omega}}{2 r^{3/2}}B_{r} -\frac{\pi^{2}}{4}B_{\varphi}+\lambda^{2} \frac{\partial}{\partial r} \left(\frac{1}{r}\frac{\partial}{\partial r} \left(rB_{\varphi}\right)\right)-\frac{\partial}{\partial r}\left(V_{r} B_{\varphi} \right).\end{equation}

   The dynamo number will change, depending on $r,$ as:
   \begin{equation}D(r)=\frac{R_{\alpha}R_{\omega}}{r^{\frac{3}{2}}}.\end{equation}
   It will decrease for larger values of the distance $r$. We can also use the effective value of this coefficient, taking it for some value $r_{0},$ for example $r_{0}=R/2$:
   \begin{equation}D_{eff}=\frac{R_{\alpha}R_{\omega}}{r_{0}^{\frac{3}{2}}}.\end{equation}

   The result of comparison of the model with the constant dynamo number ($D=9$) and the changing one ($D_{eff}=9$) is shown on Fig.2. For both cases, we take the zero conditions. We can see that in the changing dynamo number, the field in narrow inner parts of the disk is comparable to the equipartition value, but it decreases quite fast and in the outer parts, it is even smaller than in the model with constant dynamo number. It can be seen that in a small region near the inner boundary the field is a bit stronger than the equipartition value. We stress that our model of algebraic alpha-quenching admits that $\alpha$ changes sign in course of dynamo suppression. This option is not excluded, in principle, however, it is is far from being established as an obligatory property of dynamos. It is interesting to try another types of alpha-quenching, including a dynamical equation for $\alpha$ in the context of the problem~\citep{Sur07,Mikhailov13}.
    \begin{figure*}
   \centering
   \includegraphics[width=11cm]{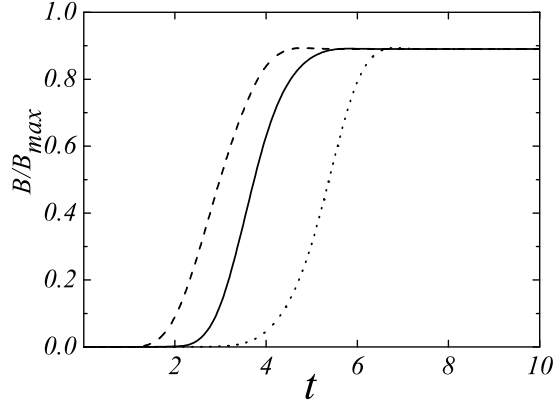}
   \caption{Time dependence of the magnetic field for different $r_{in}$ with a zero-boundary condition. Dashed line shows the case $r_{in}=0.01,$ solid line -- the case $r_{in}=0.1$ and the dotted line -- the case $r_{in}=0.2$.}
              \label{FigGam}%
    \end{figure*}

   The boundary conditions influence the magnetic field depending on the position of the inner boundary. Thus, we have tried to model the field for different $r_{in}$ values (Fig.3). It can be seen that the magnetic field strongly depends on it: for larger $r_{in}$, the field in the inner parts of the disk is smaller. However, the field for $r>0.3$ does not principally depend on it. We have also studied the time evolution of the field. We can see that for different $r_{in}$ values, we have different times for reaching the stationary value (Fig. 4).

   For a wide range of the accretion disks (e.g., near Schwarzschild black holes), we can say that the inner radius of the accretion disk is 3--5 times higher than the radius of the central object, so we take  $r_{c.o.}= 0.3 r_{in}$. However, for accretion disks around white dwarfs and neutron stars. there is such a thin boundary layer between the disk and the central object, so the two radii are approximately equal. Thus, we compared it with the case when $r_{c.o.}=r_{in}$. The result is shown on Fig.~5. Also, it is necessary to note that according to the standard accretion disk model~\citep{Shakura73},~\citep{Frank2002}, the factor $1-\sqrt{\frac{r_{c.o.}}{r}}$ in the expression for $\lambda$ is raised to the power $3/5$ ($\lambda \sim \left( 1-\sqrt{\frac{r_{c.o.}}{r}} \right)^{3/5}$).
   The result of the comparison is shown on Fig. 5  as well. We can conclude that the type of the model for the $\lambda$ coefficient and $r_{c.o.}$ does not influence the results very much and the results are robust to such changes.
   \begin{figure*}
   \centering
   \includegraphics[width=11cm]{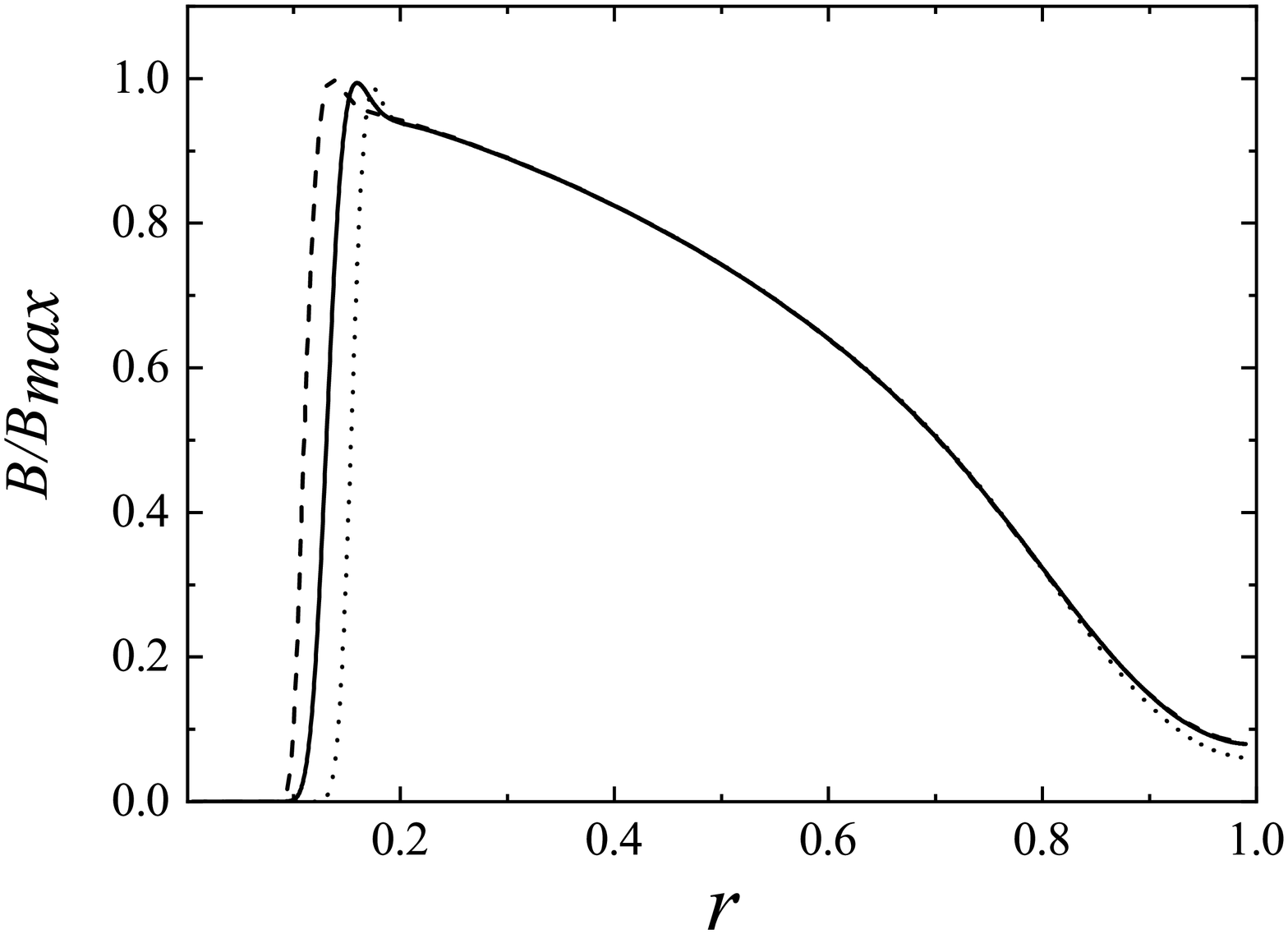}
   \caption{Dependence of the magnetic field on the distance to the center of the accretion disk for different models of the parameters for Deff=9. Solid line shows $\lambda \sim \left( 1-\sqrt{\frac{r_{c.o.}}{r}} \right)^{3/20}$ and $r_{c.o.} = r_{in};$ dashed line shows $\lambda \sim \left( 1-\sqrt{\frac{r_{c.o.}}{r}} \right)^{3/20}$ and $r_{c.o.}=0.3r_{in};$ dotted line shows $\lambda \sim \left( 1-\sqrt{\frac{r_{c.o.}}{r}} \right)^{3/5}$ and $r_{c.o.} = r_{in}$.}

    \end{figure*}

Furthermore, we study how the dynamo number influences the magnetic field values and its structure. Figure~6 shows the magnetic field for different $D$. It can be seen that large dynamo number will enlarge the magnetic field, too.

 \begin{figure*}
   \centering
   \includegraphics[width=11cm]{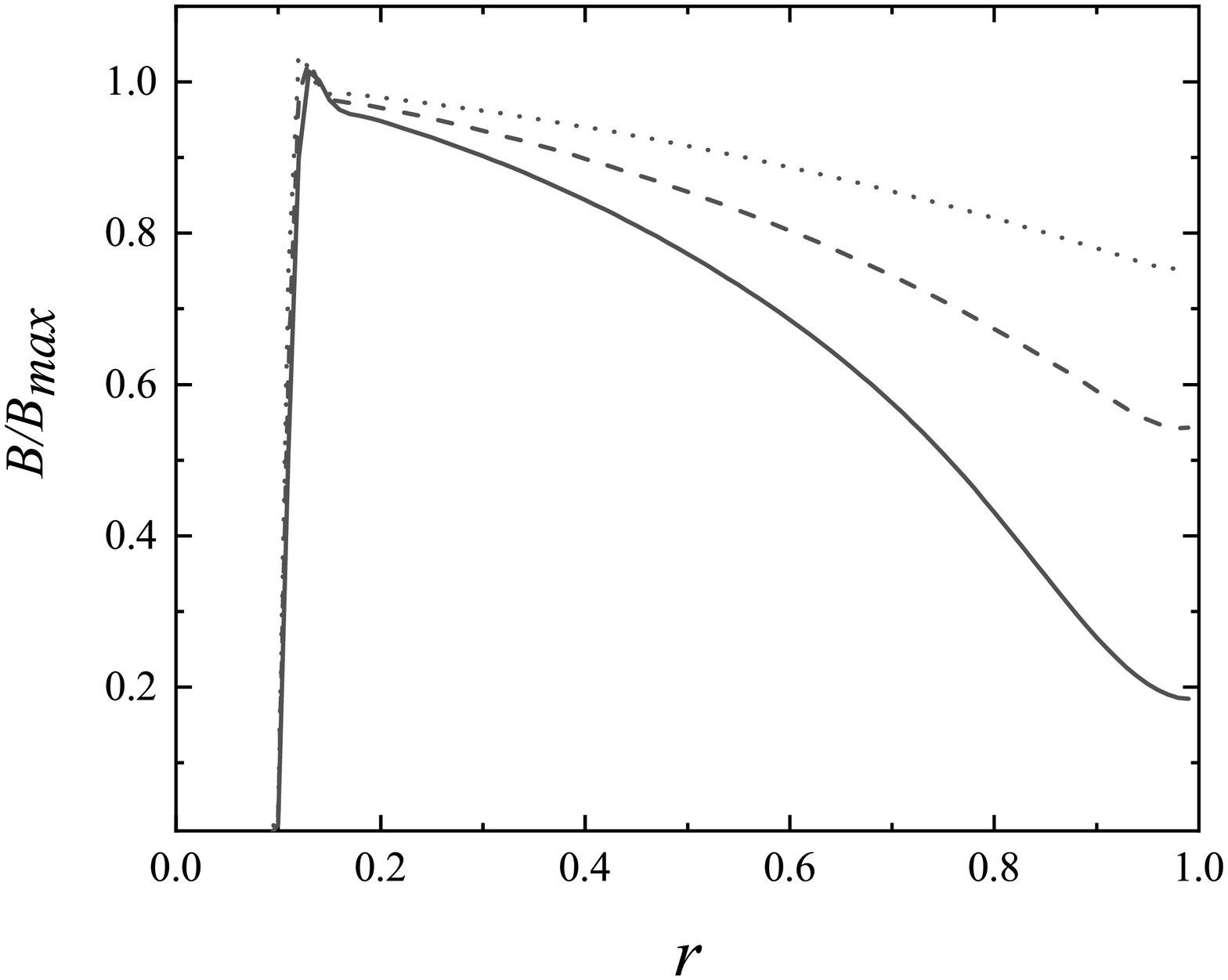}
   \caption{Dependence of the magnetic field on the distance to the center of the accretion disk for different dynamo numbers.  Solid line shows the case, when $D_{eff} = 10 $,  dashed line --  the case, when  $D_{eff} = 15$ and dotted line shows the case, when $D_{eff} = 25$.}

    \end{figure*}

   In our previous calculations, we used the simplest model for the nonlinearity. However, if we study the process of the saturation in detail~\citep{Torkelsson94}, taking into account the helicity fluxes~\citep{Shukurov05}, we should rewrite the equations by:
 \begin{equation}\frac{\partial B_{r}}{\partial t}=-\frac{R_{\alpha}B_{\varphi}}{ 1+\frac{B^{2}}{B^{*2}}} -\frac{\pi^{2}}{4}B_{r}+\lambda^{2} \frac{\partial}{\partial r} \left(\frac{1}{r}\frac{\partial}{\partial r} \left(rB_{r}\right)\right)-\frac{V_{r}}{r}\frac{\partial (r B_{r})}{\partial r},\end{equation}
   \begin{equation}\frac{\partial B_{\varphi}}{\partial t}=-\frac{3R_{\omega}B_{r}}{2r^{3/2}} -\frac{\pi^{2}}{4}B_{\varphi}+\lambda^{2} \frac{\partial}{\partial r} \left(\frac{1}{r}\frac{\partial}{\partial r} \left(rB_{\varphi}\right)\right)-\frac{\partial}{\partial r}\left(V_{r} B_{\varphi} \right).\end{equation}

   The results for this model in comparison with (\ref{eqnoz1})--(\ref{eqnoz2}) are given in Fig.~7. We can see that the results do not strongly depend on the type of nonlinearity. If $B$ is significantly smaller than $B^{*},  $ then $\left( 1- \frac{B^2}{B^{*2}}\right) \approx \left( 1+ \frac{B^2}{B^{*2}}\right)^{-1}$ and the results are nearly the same.

\begin{figure*}
   \centering
   \includegraphics[width=11cm]{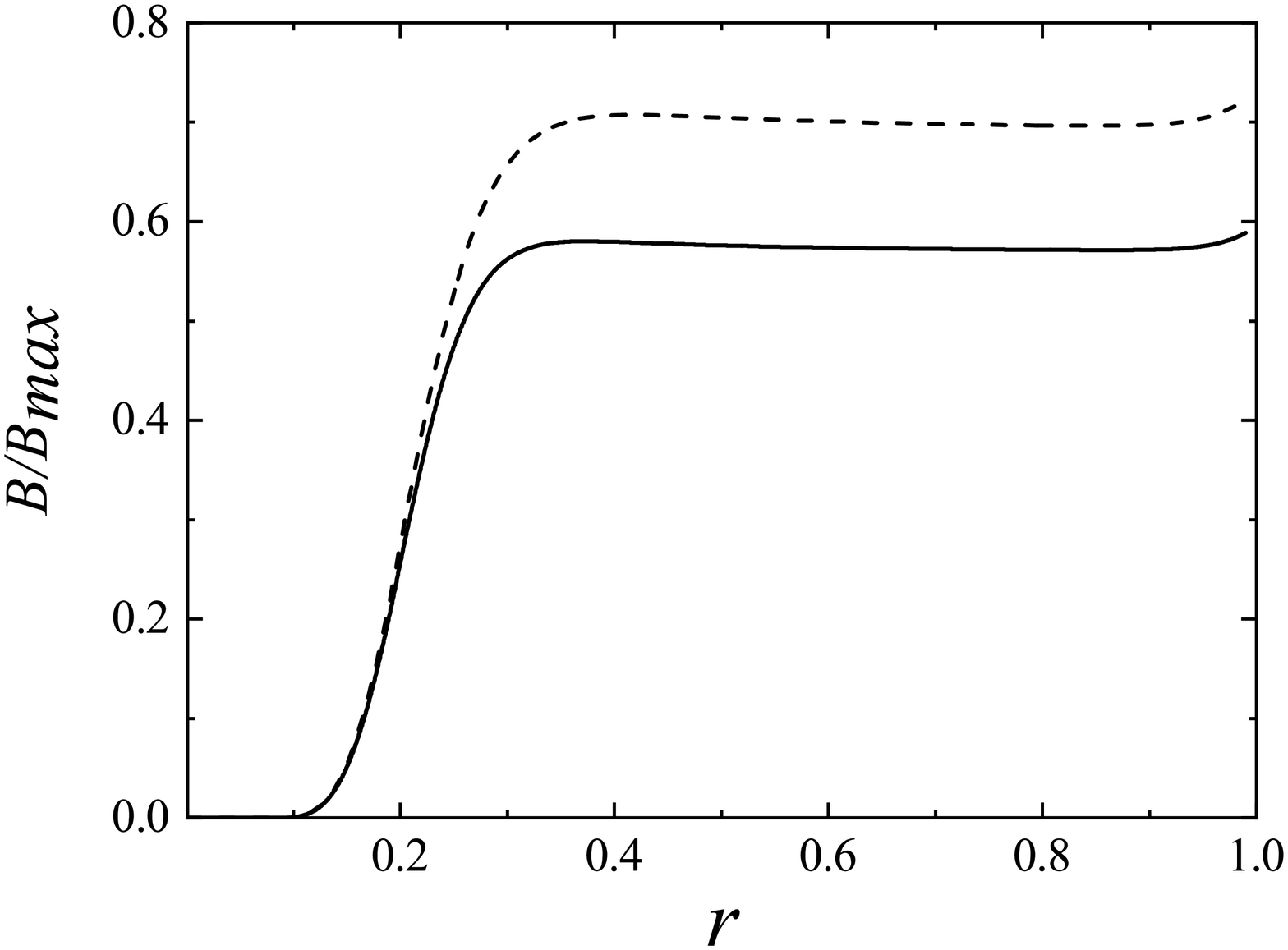}
   \caption{Dependence of the magnetic field on the distance to the center of the accretion disk for different types of nonlinearity.  Solid line shows the case where nonlinearity is $\left( 1- \frac{B^2}{B^{*2}}\right) $,  dashed line -- the case where nonlinearity is   $\left( 1+ \frac{B^2}{B^{*2}}\right)^{-1}$.}

    \end{figure*}

 Also, it is possible to take into account the dependence of the vertical dissipation and the alpha-effect on changing $h$. It can be included by using the following modifications of the equations:
$$\frac{\partial B_{r}}{\partial t}=-\frac{R_{\alpha}B_{\varphi}}{r^{9/4}}\left(1-\sqrt{r_{c.o.}/r}\right)^{-3/10} \left(1-\frac{B^{2}}{B^{*2}}\right)-$$
 \begin{equation} -\frac{\pi^{2}}{4r^{9/4}}\left(1-\sqrt{r_{c.o.}/r}\right)^{-3/10}B_{r}+\lambda^{2} \frac{\partial}{\partial r} \left(\frac{1}{r}\frac{\partial}{\partial r} \left(rB_{r}\right)\right)-\frac{V_{r}}{r}\frac{\partial (r B_{r})}{\partial r},\end{equation}
 $$\frac{\partial B_{\varphi}}{\partial t}=-\frac{3R_{\omega}B_{r}}{2r^{3/2}} -\frac{\pi^{2}}{4r^{9/4}}\left(1-\sqrt{r_{c.o.}/r}\right)^{-3/10}B_{\varphi}+$$
   \begin{equation}+\lambda^{2} \frac{\partial}{\partial r} \left(\frac{1}{r}\frac{\partial}{\partial r} \left(rB_{\varphi}\right)\right)-\frac{\partial}{\partial r}\left(V_{r} B_{\varphi} \right).\end{equation}
We can see (Fig. 8) that for this case, the critical dynamo number is enlarged and the field is generated only in the outer parts of the accretion disk. It is interesting to study magnetic field growth for larger values of the dynamo number, which correspond to realistic cases based on the physics of accretion disks. For example, Torkelsson~\&~Brandenburg (1994b) describe the magnetic field using parameters $C_{\alpha}=\frac{\alpha_{0}R}{\nu}$ and $C_{\Omega}=\frac{\Omega_{0}R^{2}}{\nu}$. In our model we have $D\sim\left(\frac{h}{R}\right)^3 C_{\alpha}C_{\Omega}$. Thus, the values corresponding to the values used by~\citep{Torkelsson94} are expected to be on  the order of $10^{2...3}$. We also analyze the time dependence of the field (Fig. ~9). We can see that for $D=10^{3}$, the field grows quite rapidly (so, the figure uses the logarithmic scale). There is a slight difference in the results, which have been obtained by different approaches. This means that more realistic models (26), (27) give smaller growth rates of the magnetic field. As for intensive dynamo action, the growth rate enlarges with the dynamo number proportionally, according to $D^{1/2}$.

\begin{figure*}
   \centering
   \includegraphics[width=11cm]{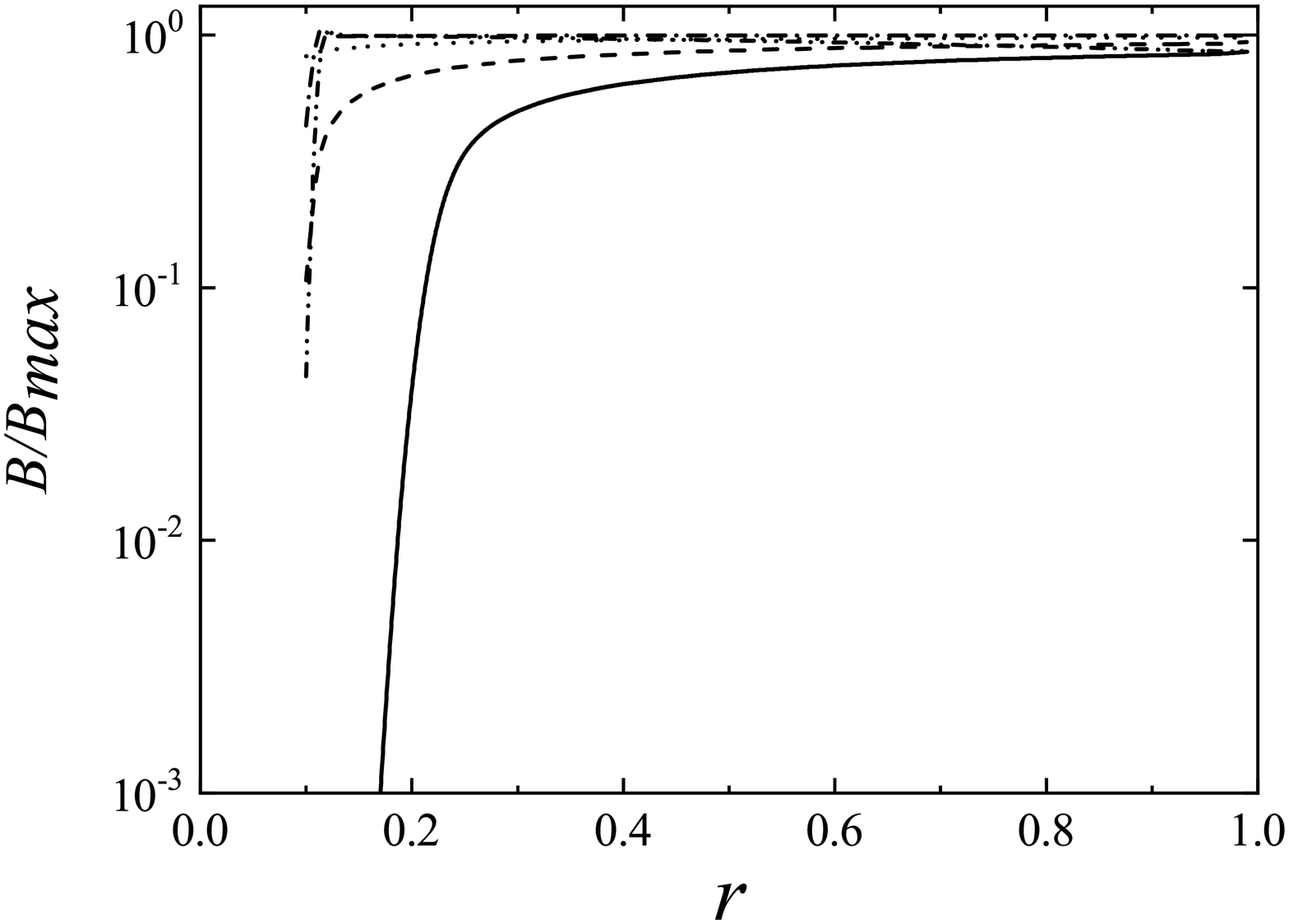}
   \caption{Radial dependence of the magnetic field on the distance to the center of the accretion disk for different cases. According to Eqs. (26), (27): the solid line shows the case where $D=15$, dashed line -- the case where $D=30$, dotted line -- the case where $D=100$, dot-dashed line -- the case where $D=1000$. According to Eqs. (8), (9): dot-dot-dashed line -- the case where $D=15$.}

    \end{figure*}

\begin{figure*}
   \centering
   \includegraphics[width=11cm]{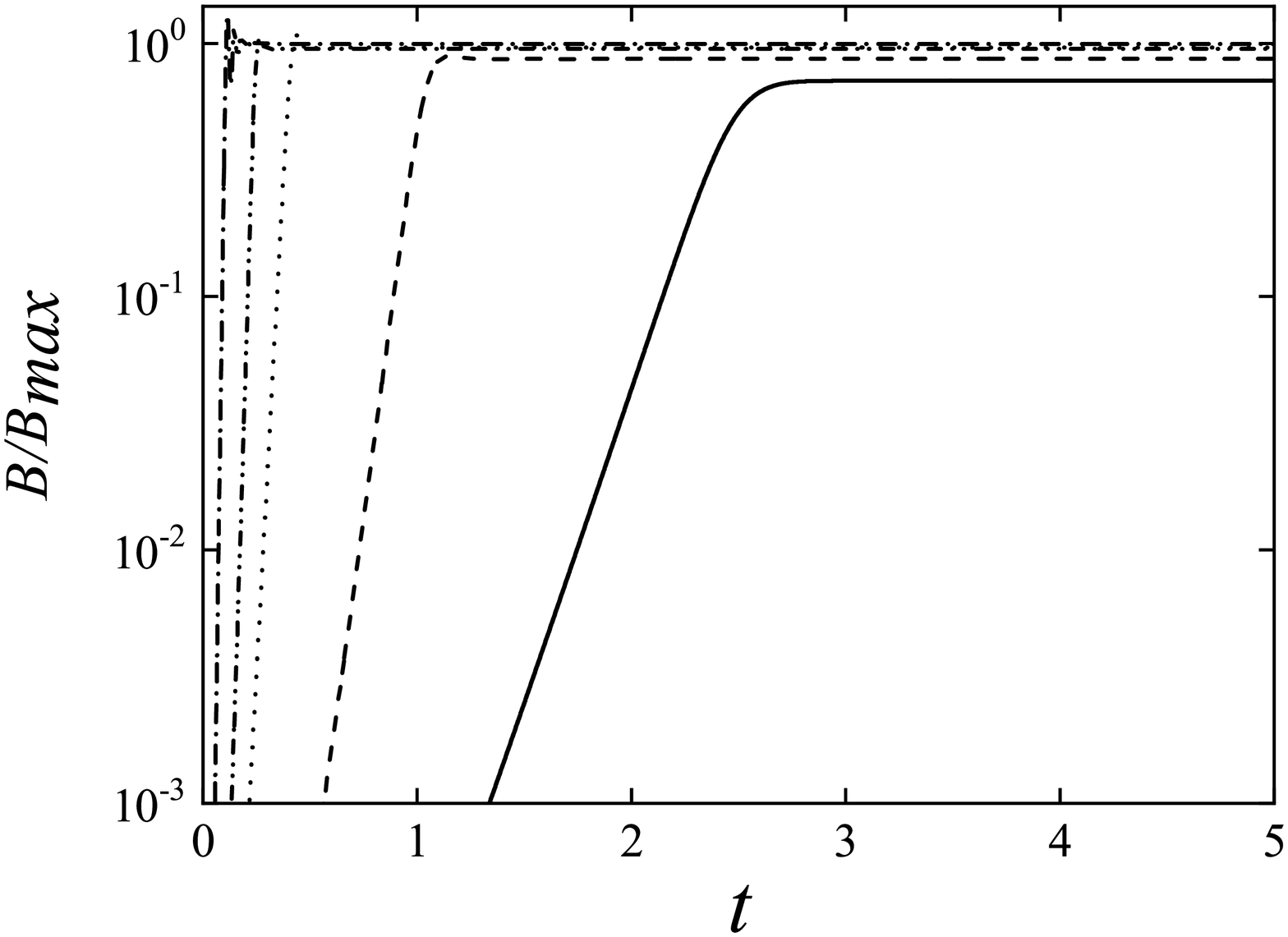}
   \caption{Time dependence of the magnetic field on the distance to the center of the accretion disk for different cases. According to Eqs. (26), (27): solid line shows the case where $D=15$, dashed line -- the case where $D=30$, dotted line -- the case where $D=100$, dot-dashed line -- the case, when $D=1000$. According to Eqs. (8), (9): dot-dot-dashed line -- the case where $D=15$.}

    \end{figure*}

   \section{Conclusion}

   In this work, we show that the magnetic field in the accretion disks surrounding different massive bodies can grow using the dynamo mechanism. To model it, we used the no-$z$ approximation, which has been developed to study the field in the accretion disks. Magnetic field models with constant dynamo number and the changing one have been developed. For both, it has been shown that the boundary conditions play a principal role in the field evolution and structure. If we take the boundary conditions with zero field on the inner boundary, the field is much smaller than the equipartition value. It means that the magnetic field is less than the equipartition value, so we can neglect most of the effects connected to the influence of the magnetic field on the motions of the medium. In our view, this model can be used to describe the magnetic field in different cases related to accretion disks of different nature.

  The main finding of our paper is that the dynamo-generated magnetic field configuration in accretion disks is quite similar to the one known for galactic disks. Of course, an appropriate rescaling of spatial scale and another governing parameters is presumed here, as well as a physically motivated boundary condition at the inner boundary of the disk. In this context, the main difference between accretion disks dynamo and galactic dynamo can be presented as follows. Dynamo action in the galactic disk is more or less independent of magnetic field evolution in the nearby galactic center. This is why it seems appealing at first glance to presume a condition in a particular galactic dynamo model to separate magnetic field evolution in the disk from that of the nearby galactic center. However, as accretion disk contains accretion flows, dynamo modeling in the disk should be separated more carefully from physical processes in the nearby corresponding central body. Using a naive inner boundary condition, which fully isolates magnetic configuration in the disk from what happens nearby the central body, we can obtain a non-physical magnetic configuration with a super-equipartition magnetic field strength. Using a more or less realistic inner boundary condition (at least for
modest dynamo numbers) such as zero field ones, we can obtain a configuration with magnetic field strength that is comfortably weaker than the equipartition.

    \section{Acknowledgements}

   Authors are grateful to N.I.Shakura for his advice and useful comments.
   The work is partially supported by Russian Ministry of Science and Higher Education, agreement No. 075-15-2019-1621.
   This work of M.V.P. was supported by Theoretical Physics and Mathematics Advancement Foundation "BASIS" under grant 20-2-1-17-1.
   This work is partially supported by the grant: Binary stars with compact objects, K$\Pi$-06-H28/2 08.12.2018
(Bulgarian National Science Fund).
   Authors thank the referee for his review and the questions, which allowed to improve the work.

\end{document}